\documentclass[fleqn,usenatbib]{mnras}

\usepackage{newtxtext}
\usepackage[slantedGreek]{newtxmath}

\usepackage[T1]{fontenc}
\usepackage{ae,aecompl}

\usepackage{enumerate}
\usepackage{graphicx}	
\usepackage{amsmath}	
\usepackage{amssymb}	
\usepackage{natbib}
\usepackage{longtable,lscape}

\bibpunct{(}{)}{;}{a}{}{,}
\title[Cluster membership for SV~Vul]{Cluster membership for the long period Cepheid calibrator SV~Vul\thanks{Partially based on observations made with the Nordic Optical Telescope.}}

\author[I. Negueruela et al.]{
I. Negueruela,$^{1}$\thanks{E-mail: ignacio.negueruela@ua.es}
R. Dorda,$^{2}$ and A. Marco,$^{3}$
\\
$^{1}$Departamento de F\'{\i}sica Aplicada, Facultad de Ciencias, Universidad de Alicante,\\ Carretera de San Vicente s/n, E03690, San Vicente del Raspeig, Spain\\
$^{2}$Instituto de Astrof\'{\i}sica de Canarias,  V\'{\i}a L\'actea s/n, E38200, La Laguna, Tenerife, Spain\\
$^{3}$Departamento de F\'{\i}sica, Ingenier\'{\i}a de Sistemas y
Teor\'{\i}a de la Se\~{n}al, Universidad de Alicante,\\ Carretera de San Vicente s/n,
E03690, San Vicente del Raspeig, Spain\\
}

\date{Accepted 2020 March 22. Received 2020 March 22; in original form 2020 January 16}

\pubyear{2019}

\begin{document}
\label{firstpage}
\pagerange{\pageref{firstpage}--\pageref{lastpage}}
\maketitle

\begin{abstract}
Classical Cepheids represent the first step of the distance scale ladder. Claims of tension between the locally calculated Hubble constant and the values deduced from \textit{Planck}'s results have sparked new interest in these distance calibrators. Cluster membership provides an independent distance measurement, as well as astrophysical context for studies of their stellar properties. Here we report the discovery of a young open cluster in the vicinity of SV~Vul, one of the most luminous Cephedis known in the Milky Way. \textit{Gaia} DR2 data show that SV~Vul is a clear astrometric and photometric member of the new cluster, which we name Alicante~13. Although dispersed, Alicante~13 is moderately well populated, and contains three other luminous stars, one early-A bright giant and two low-luminosity red supergiants. The cluster is about 30~Ma old at a nominal distance of 2.5~kpc. With this age, SV~Vul should have a mass around $10\:$M$_{\sun}$, in good accordance with its luminosity, close to the highest luminosity for Cepheids allowed by recent stellar models.
\end{abstract}

\begin{keywords}
stars: evolution -- supergiants  --
 Hertzsprung-Russell and colour-magnitude diagrams --
 stars: variables: Cepheids
 -- open clusters and associations: individual: Alicante~13 
\end{keywords}



\section{Introduction}

Cepheid variables were the earliest standard candles in use and still today constitute the first step in the cosmic distance ladder \citep[and references therein]{feast99}. With the release of \textit{Gaia} DR2 \citep{brown18} geometrical parallaxes,  their role as benchmark distance indicators through the use of Leavitt Law \citep[e.g.][]{freedmad10} is becoming even more paramount. Claims for a statistically significant difference between values of the Hubble constant derived locally and from a combination of Cosmic Microwave Background measurements and the standard cosmological models \citep[e.g.][]{riess18b} call for high-precision distance determinations. In this context, Cepheids in open clusters are doubly valuable. The cluster offers an independent way to measure the distance and provides an astrophysical context for the Cepheid. Thanks to this context, the physical properties and evolutionary stage of the Cepheid can be constrained. 

The number of Cepheids believed to belong to clusters or OB associations has increased in recent years \citep{anderson13,chen15}, but not all of them are well studied. The best characterised stars are included in period-luminosity calibrations. Among them, SV~Vul, with a 45~d pulsation period, is one of the brightest. It is indeed the star with the longest period in a number of calibrations based on Milky Way Cepheids \citep[e.g.][]{tammann03, fouque07}. In others \citep[e.g.][]{storm11}, it is superseded by the confusingly named 63~d Cepheid S~Vul (which is located only 40 arcmin away), whose geometric parallax was recently measured with the HST \citep{riess18a}.

By the use of different standard methods, the distance modulus to SV~Vul has traditionally been calculated around $\mu=11.7$~mag ($d=2.1\:$kpc; $\pi=0.46$; e.g. \citealt{fouque07}), implying an absolute magnitude approaching $M_{V}=-6$. \citet{storm11} proposed a lower value $\mu=11.4$, based on the near-IR surface brightness method. Conversely, \citet{madore17} suggest a higher value $\mu=12.0$~mag ($\pi=0.40$~mas) by applying a novel method to treat reddening. This latter value is in good agreement with the \textit{Gaia} DR2 parallax $\pi=0.37\pm0.03$~mas, although DR2 values for such a bright ($G=6.9$) star should be taken with care.

SV~Vul has long been assumed a member of the Vul~OB1 association, following \citet{turner84}. The core of Vul~OB1 is the very young (star-forming) cluster NGC~6823, which contains a few O-type stars, including the O6.5\,V((f)) HD~344784. There are, however, older stars belonging to the association, and SV~Vul has been connected to the rather older cluster NGC~6834, which lies about 2 degrees North of the Cepheid \citep{turner76}. With an estimated age of $80\:$Ma \citep{paunzen06}, NGC~6834 seems too old to be related to SV~Vul. In addition, its \textit{Gaia} DR2 parallax $\pi=0.27$ \citep{cantat18} places it too far away to belong to Vul~OB1. Moreover membership in an association for isolated stars in this direction is difficult to assess, as three distinct OB associations, Vel~OB1, Vel~OB2 and Vel~OB4, are believed to be projected one of top of the other at different distances (we will come back to this issue in the Discussion).

In this work, we demonstrate the presence in this region of a newly recognised open cluster  that includes SV~Vul as a halo member. The new cluster, which we name Alicante~13\footnote{Information about other clusters in the series can be found at {\tt https://astro.ua.es/alicanteclusters/}, where all relevant references are given.}, is not apparent to the eye, except as a compact group of three bright stars, namely HD~339063, HD~339064 and BD~$+27^{\circ}$3542, whose \textit{Gaia} DR2 astrometric parameters (listed in Table~\ref{sgs}) are fully compatible with those of SV~Vul, i.e. pmRA =$-2.14\:$mas/yr, pmDec =$-5.82\:$mas/yr, $\pi=0.37\:$mas. This group is located about 9 arcmin to the East from SV~Vul. In the following sections we identify the cluster as a concentration of early-type stars, determine its astrophysical parameters and characterise its brighter members.
 
\section{Cluster definition}

To identify the cluster, we used the Virtual Observatory tool Clusterix 2.0 \citep{balaguer20} working on \textit{Gaia} DR2 data. Clusterix is an interactive web-based application that calculates the grouping probability of a list of objects by using proper motions and the non-parametric method proposed by \citet{cabreracano90} and described in \citet{galadi98}. In its current version, Clusterix works only in the proper motion plane, ignoring all other information. For ease of computation, we restricted ourselves to DR2 objects with $G\leq16$ and errors in proper motion below $3\:$mas/yr. We selected a circle of radius 30 arcmin around the position of HD~339063 and run the tool assuming a cluster radius of 3.6 arcmin (this simply informs the tool of where it should find a significant number of objects belonging to a distinct population). Clusterix identifies over 6300 sources brighter than $G=16$ and passing its quality criteria.  Clusterix then assigns each object a probability of belonging to a distinct population rather than the field population, using an empirical determination of the frequency functions in the vector point diagram \citep{sanders71}. After this, we plotted the parallax vs. probability plane, and found a clear overdensity of objects with high probability and similar values of parallax. 

   \begin{figure}
   \centering
\resizebox{\columnwidth}{!}{\includegraphics[,clip]{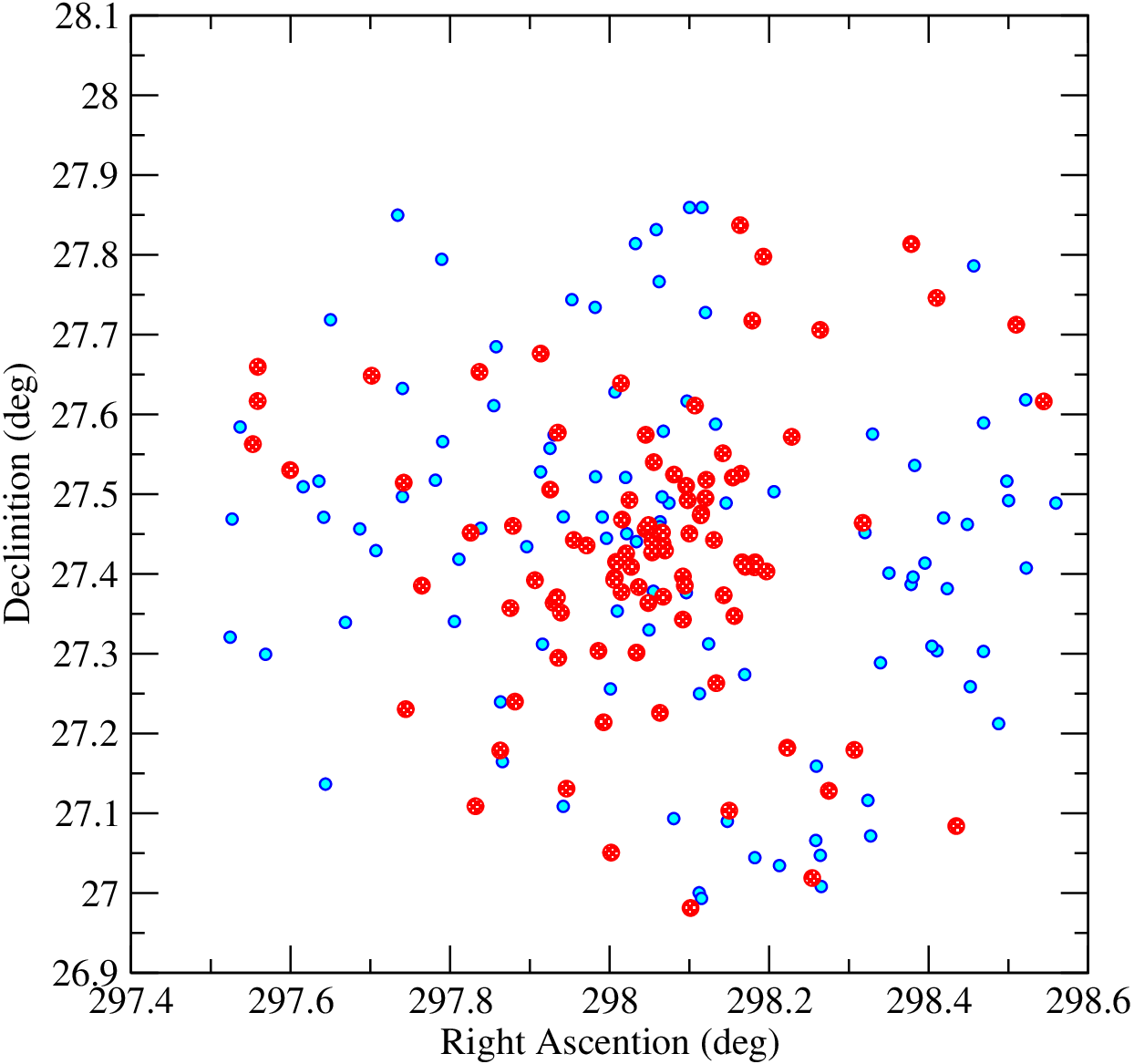}}
   \caption{Spatial distribution of the population selected by Clusterix. Small blue circles are objects whose three astrometric parameters lie within 2\,$\sigma$ of the average values. Larger red circles are objects whose parameters are less than one $\sigma$ away from these central values.
  \label{spat}}
    \end{figure}

The population thus identified is strongly concentrated in the proper motion plane and contains the four stars mentioned above. After removal of a handful of outliers, we calculated the weighted means for the astrometric parameters of the population, which turn out to be pmRA =$-2.08\pm0.21\:$mas/yr, pmDec =$-5.92\pm0.15\:$mas/yr, $\pi=0.37\pm0.06\:$mas, where the errors quoted are the standard deviations for the whole population selected. The median values of these parameters are pmRA =$-2.08$ mas/yr, pmDec =$-5.87$ mas/yr, $\pi=0.38$~mas. Their modes are  pmRA =$-2.09$ mas/yr, pmDec =$-5.87$ mas/yr, $\pi=0.37$~mas. Coincidence of mean, median and mode indicates that the central values are well defined and a more sophisticated analysis is not needed. An iterative process of 2-$\sigma$ outlier clipping leads to essentially identical median values, confirming that the results of Clusterix are statistically solid. In Fig.~\ref{spat}, we plot the spatial distribution of the objects finally selected (around 200 sources). Although the population is spread over the whole field, there is a very strong concentration towards the position of the three bright stars mentioned above.

\begin{figure*}
   \centering
\resizebox{\columnwidth}{!}{\includegraphics[,clip]{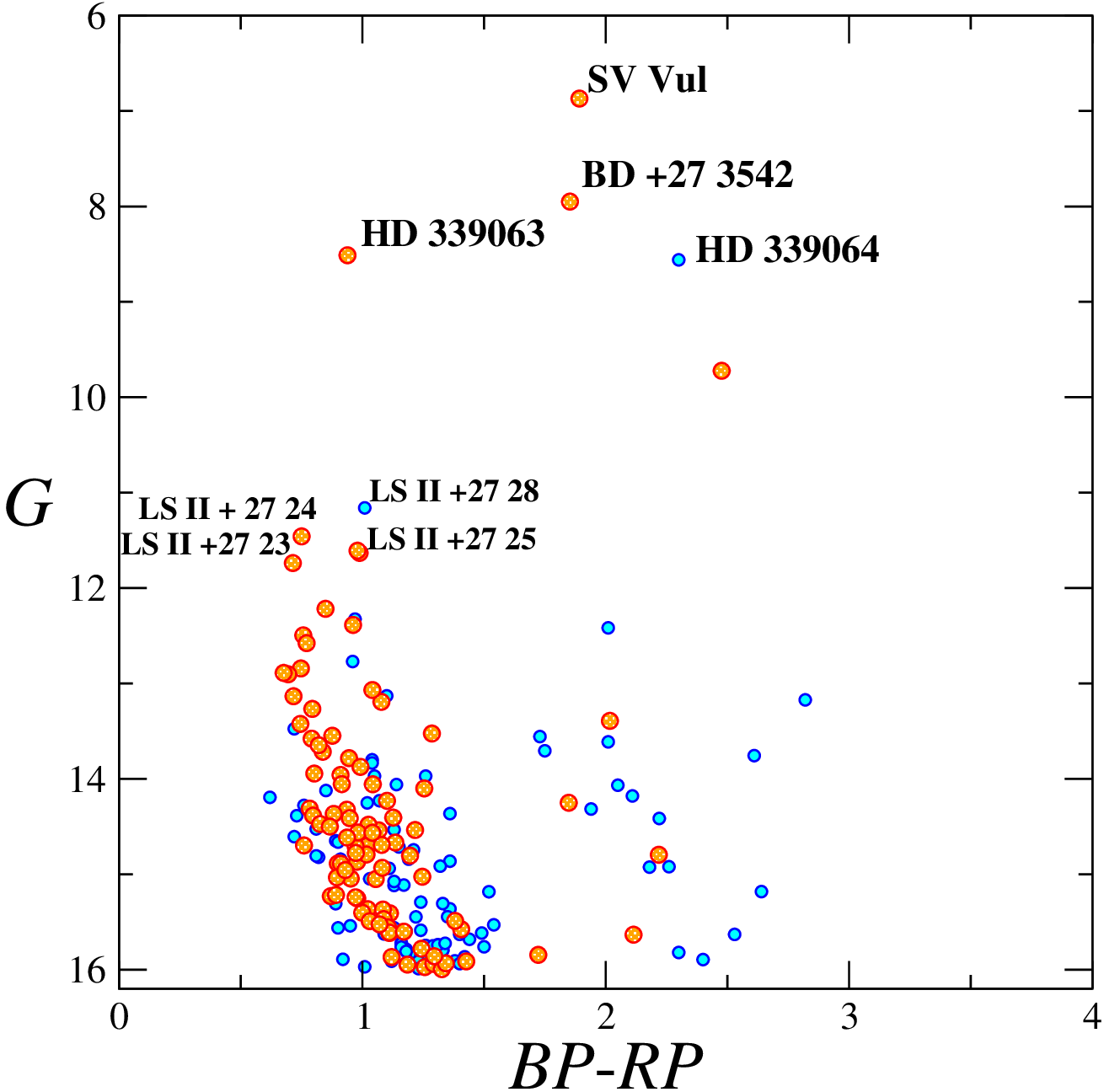}}
\resizebox{\columnwidth}{!}{\includegraphics[,clip]{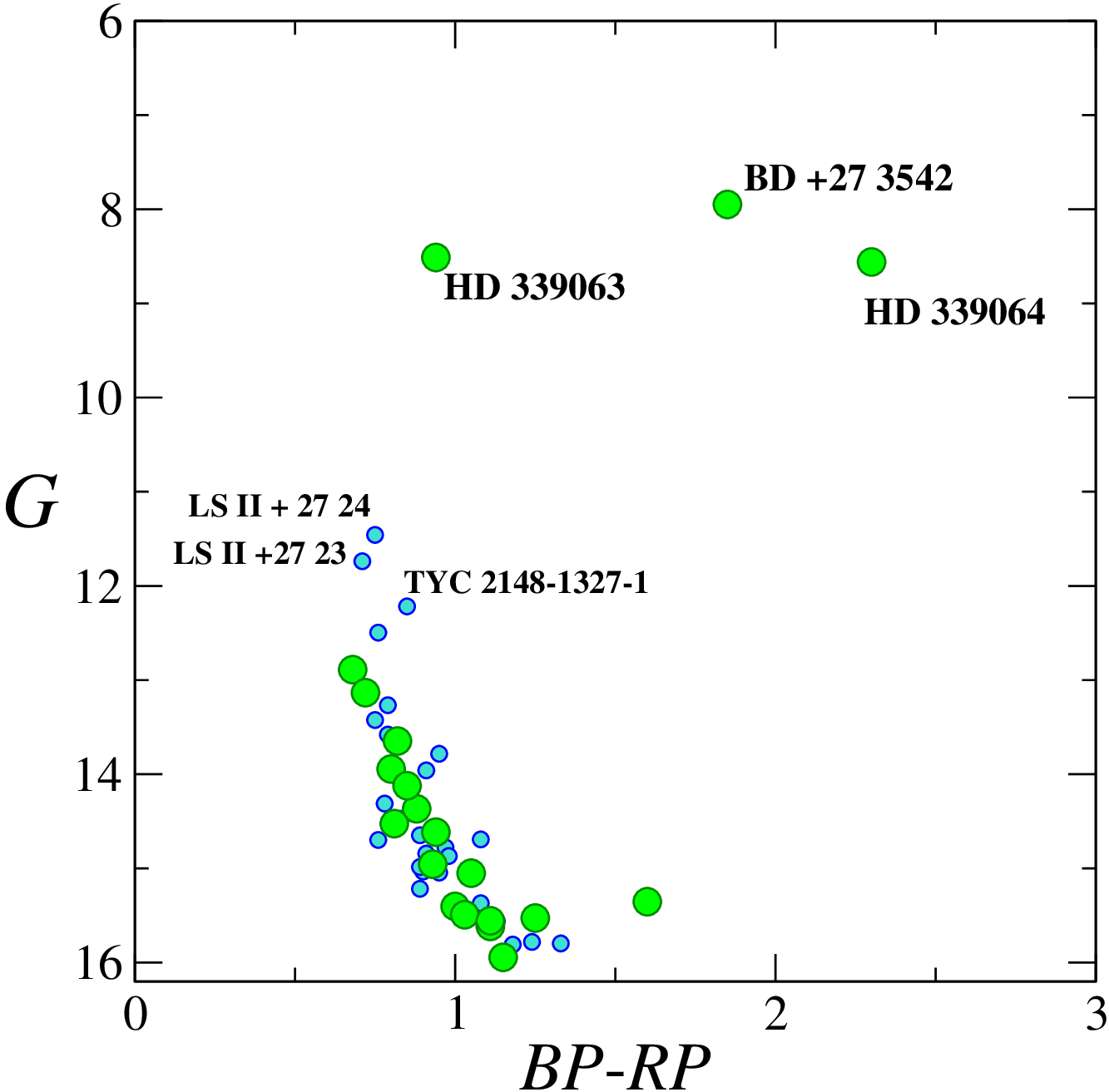}}
   \caption{\textit{Gaia} CMD for the sample selected using astrometric data. \textbf{Left panel: }The large (red) circles have astrometric paramenters within one $\sigma$ of the average values for the population selected by Clusterix. The small blue circles represent stars within 2\,$\sigma$ of the same values. \textbf{Right panel: } Only stars within 6 arcmin of HD~339063 are shown. The large green circles are stars within 3 arcmin. In both diagrams the brightest stars are labeled with their catalogue names.\label{gaiaphot}}
    \end{figure*}

To quantify this, we used the Automated Stellar Cluster Analysis (ASteCA) code \citep{perren}.  We run ASteCA on the selection shown in Fig.~\ref{spat}, i.e. stars whose three astrometric parameters fall within 2\,$\sigma$ of the average values for the population. A King profile does not provide a good fit, as the central  distribution is very asymmetric, but a strong overdensity is detected, with about 40 stars above the background value within $\sim5$~arcmin of the central position given. This is however, an underestimate of the cluster contrast, since we are only utilising stars with the same astrometric parameters and thus the "background" sources should be mostly the dispersed population of the cluster halo. In the left panel of Fig.~\ref{gaiaphot}, we plot the \textit{Gaia} photometry for these stars. We can see that around 20 stars (i.e., $\sim10$\% of the objects) occupy positions in the \textit{Gaia} CMD suggesting that they are interlopers. All the other stars seem to belong to a single population. The spread in ($BP$-$RP$) is very likely due to differential reddening. For instance, all the bright stars ($G\approx11$\,--\,12) around ($BP$-$RP$)$\approx1.0$ lie to the East of the cluster, mostly at large distances ($>15\arcmin$). All the objects located at the top of the main stellar sequence are catalogued early-type stars. They include LS\,II~$+27^{\circ}$23, 24, 25 \& 28, TYC~2148-1327-1, and the emission-line star Hen~3-1788.

The cluster is small and disperses into the surrounding association. Of the $\sim180$ likely members of the association shown in the left panel of Fig.~\ref{gaiaphot}, there are 47 stars within 6 arcmin of HD~339064 and 90 stars within 12 arcmin. The right panel of Fig.~\ref{gaiaphot} shows the population within 6~arcmin. Differential reddening is reduced to a minimum and the sequence is now much better defined. Only one of the faintest stars is located at some distance from the photometric sequence.

\section{Observations and analysis}

High-resolution spectra of the three bright stars and SV~Vul itself were taken with the FIbre-fed Echelle Spectrograph (FIES) attached to the 2.56~m Nordic Optical Telescope (NOT; La Palma, Spain) in service mode during the night of 2019, June 4. FIES is a cross-dispersed high-resolution echelle spectrograph, mounted in a heavily insulated building separated from and adjacent to the NOT dome, with a maximum resolving power $R=67\,000$. The entire spectral range 370\,--\,830 nm is covered without gaps in a single, fixed setting. Extended coverage up to 900 nm is available with minor inter-order wavelength gaps. In the present study, we used the low-resolution mode with $R=25\,000$. The spectra were homogeneously reduced using the
FIEStool\footnote{http://www.not.iac.es/instruments/fies/fiestool/FIEStool.html} software in advanced mode. Using a complete set of bias, flat, and ThAr arc frames, the FIEStool pipeline provides wavelength calibrated, blaze corrected, order merged spectra.

Classification spectroscopy of blue stars in the area was obtained in service mode during the night of 2019, June 10, with the imager and spectrograph ALFOSC attached to the NOT. We used the grism \#18 combined with a $1\arcsec$ slit to obtain intermediate resolution spectroscopy. Grism \#18  covers the 3450\,--\,5350\:\AA{} range with a nominal dispersion of 0.9\,\AA/pixel. The resolving power for this configuration is $R\sim1000$. These data were reduced following standard procedures with reduction software inside the {\em Starlink}
suite \citep{currie14}. 

Spectral classification of the early-type stars has been carried out by comparison to a grid of standards \citep{negueruela19} degraded to the same resolution and application of classical criteria. Spectral classification of late-type supergiants has been carried out by following the procedure specified in \citet{dorda18} on their spectra degraded to classification resolution.

Radial velocities (RVs) were calculated for stars with high-resolution spectra by using iSpec\footnote{A free integrated spectroscopic software powered by python, available at {\tt https://www.blancocuaresma.com/s/iSpec}} \citep{ispec14,ispec19}. As a first step, telluric features were removed to improve the radial velocity determination. Then we cross-match correlated each of our spectra against a high-resolution linelist mask of the Sun (provided by iSpec) covering our whole spectral range, and we obtained the corresponding mean velocities by fitting a second order polynomial near the peak to the velocity profile. The errors were automatically calculated by iSpec following the procedures described in \cite{zucker03}. The RVs measured are shown in Table~\ref{sgs}, where they are compared to the DR2 values (note that HD~339063, being a blue star has no DR2 RV).

\section{Results}

\subsection{Stellar content}
\label{content}

\begin{figure}
   \centering
\resizebox{\columnwidth}{!}{\includegraphics[angle=-90,clip]{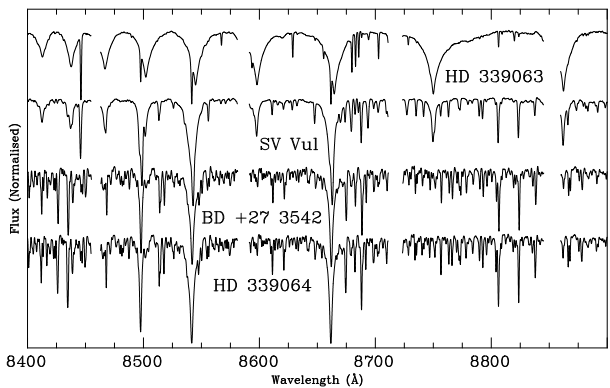}}
   \caption{Spectral sequence of luminous stars around the region covered by the \textit{Gaia} RVS. The small gaps are due to the lack of order overlap in FIES for the extreme red. 
  \label{sequence}}
    \end{figure}

There are four luminous stars that appear as astrometric and photometric members of the cluster. Their derived parameters are listed in Table~\ref{sgs}. The brightest among them is the well-studied Cepheid SV~Vul, located in the cluster halo. Spectra of the four stars in the \textit{Gaia} RVS spectral region are shown in Fig.~\ref{sequence}. SV~Vul appears as an F8\,Iab supergiant. Its spectral type is known to vary between F7 and K0 \citep{code47}. Its radial velocity oscillates periodically between $\approx+20\:$km\,s$^{-1}$ and  $\approx-25\:$km\,s$^{-1}$ with a systemic velocity\footnote{The systemic velocity cannot be determined with high accuracy because the oscillation period is not stable \citep[e.g.][]{bersier94}.} $\approx-1\:$km\,s$^{-1}$ \citep{bersier94}. The high dispersion of the \textit{Gaia} RV measurements testifies to this variability. Our measurement is typical of the velocity seen at early spectral types.

\begin{figure}
   \centering
\resizebox{\columnwidth}{!}{\includegraphics[angle=-90,clip]{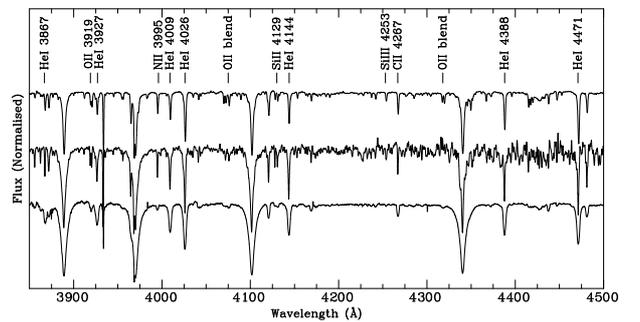}}
   \caption{The blue component of BD~$+27^{\circ}$3542 (middle) compared to stars of similar spectral types. Contamination by the K-type brighter companion is evident beyond 4\,200\,\AA. The top spectrum is 3~Gem (HD~42087; B3\,Ib). The bottom spectrum is HD~21483 (B3\,III).
  \label{blueguy}}
    \end{figure}

The position of BD~$+27^{\circ}$3542 in the \textit{Gaia} CMD suggests a yellow star, similarly to SV~Vul. However, its spectrum shows it to be a composite, containing a very luminous red star and a B-type object. The red star, for which we derive a K3\,Iab spectral type, completely dominates the red spectrum (see Fig.~\ref{sequence}). The blue star can be seen up to $\sim 5\,000\,$\AA, and dominates the spectrum bluewards of $\sim 4\,500\,$\AA. We display this region in Fig.~\ref{blueguy}, together with comparison spectra. The lines of the blue component are very narrow, indicating a very low projected rotational velocity. The wings of the Balmer lines can be used to estimate its luminosity, which turns out higher than that of the B3\,III standard HD~21483. We estimate a spectral type B3\,II--III, with some uncertainty due to the strong contamination by the red star. Given that \textit{Gaia} does not separate the two stars, they are very likely to form a physical binary. Against this, our RV measurement is not significantly different from the \textit{Gaia} value, which is the median of observations taken between 2014 and 2016, i.e. more than three years before our observation. On the other hand, a chance projection is highly unlikely, especially if we consider that the blue component is the second most luminous blue star in the cluster.

\begin{table*}
        \centering
        \caption{Parameters for the stars observed at high resolution. Spectral type and a measurement of RV are from our spectra, while the other parameters are from \textit{Gaia} DR2.}
        \label{sgs}
        \begin{tabular}{lcccccccccc}
                \hline
                \noalign{\smallskip}
Star&  Spectral & pm (RA) & pm (Dec)& $\pi$  & $G$ & $BP$-$RP$ &RV (\textit{Gaia}) & RV \\ 
&type & (mas) & (mas) & (mas) & (mag) & (mag)& (km\,s$^{-1}$) & (km\,s$^{-1}$)  \\
\noalign{\smallskip}
\hline
\hline
\noalign{\smallskip}%
SV~Vul   &F8\,Iab & $-2.14\pm0.05$ &   $-5.82\pm0.05$ &  $0.37\pm0.03$& 6.87 & 1.89 & $+3.2\pm4.3$& $-14.5\pm0.3$ \\
HD~339063   & A2\,II &        $-2.22\pm0.05$ &   $-5.89\pm0.05$ &  $0.41\pm0.03$& 8.51 & 0.94 & $-$ & $-1.5\pm0.6$  \\
HD~339064  &K1\,Ib&     $-2.24\pm0.05$ &   $-5.72\pm0.06$ &  $0.35\pm0.03$&  8.56 & 2.30&  $+6.58\pm0.39$& $-3.5\pm0.1$\\
BD~$+27^{\circ}$3542&B3\,II + K3\,Iab&$-2.24\pm0.05$ &   $-5.92\pm0.05$ &  $0.37\pm0.03$&  7.95 & 1.85 & $-0.14\pm0.25$ &$-0.5\pm0.2$ \\

\noalign{\smallskip}
                \hline
        \end{tabular}
\end{table*}

\begin{figure}
   \centering
\resizebox{\columnwidth}{!}{\includegraphics[angle=-90,clip]{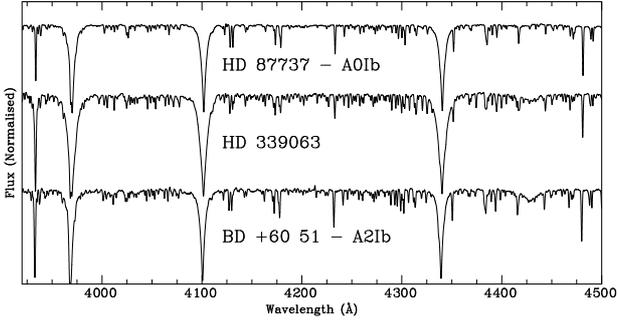}}
   \caption{Classification spectrum of HD~339063, with two comparison stars. HD~87737 is a primary MK standard, while BD~$+60^{\circ}$51 was characterised by \citet{verdugo99}.  
  \label{bluesgs}}
    \end{figure}

As can be seen in  Fig.~\ref{gaiaphot}, HD~339063 is a blue star. Its spectrum in the \textit{Gaia} RVS spectral region is shown in Fig.~\ref{sequence}, while its classification spectrum is presented in Fig.~\ref{bluesgs}, together with comparison spectra. The strong metallic spectrum and lack of \ion{He}{i} lines imply an A spectral type. Its lines are somewhat broader than those of the Ib stars displayed as comparison, suggesting a lower luminosity. The main temperature diagnostic, the ratio of \ion{Ca}{ii} K to H$\epsilon$, suggests that HD~339063 is not much later than A2. Comparison to BD~$+60^{\circ}$51 shows a slightly stronger and more complex metallic spectrum, except for those lines that are sensitive to luminosity, such as \ion{Fe}{ii}\,4233\AA, the \ion{Fe}{ii} and  \ion{Ti}{ii} blend at 4172\,--\,8\AA, and the  \ion{Si}{ii} doublet at 4128\,--\,30\AA. In view of this, we classify HD~339063 as A3\,II. Our RV measurement, $-1.5\pm0.6\:$km\,s$^{-1}$, is fully compatible with the systemic velocity of SV~Vul and BD~$+27^{\circ}$3542.

Finally, HD~339064 is a slightly less luminous red star, for which we estimate a spectral type K1\,Ib. Our radial velocity measurement, $-3.5\pm0.1\:$km\,s$^{-1}$, is very different from the \textit{Gaia} value and much more in line with the velocities of the other members. We do not find evidence for a second stellar component, while the dispersion of the \textit{Gaia} DR2 RVs does not suggest high variability. The reasons for this discrepancy are thus unclear.

\begin{figure}
   \centering
\resizebox{\columnwidth}{!}{\includegraphics[angle=-90,clip]{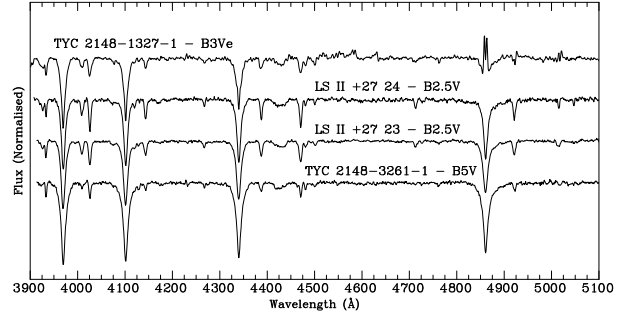}}
   \caption{Classification spectra of stars selected as upper main-sequence cluster members.
  \label{bluestars}}
    \end{figure}

We also obtained classification spectra of the brightest stars of the main sequence in the photometric CMD. These spectra are displayed in Fig.~\ref{bluestars} and their main parameters are listed in Table~\ref{tabblues}. The two brightest members are the catalogued early-type objects LS\,II~$+27^{\circ}$23 \& 24. Their spectra are very similar, and we classify both as B2.5\,V. A slightly fainter member is TYC~2148-1327-1, catalogued as an emission line star (HBHA~2703-20). It has features typical of a Be star, with double-peaked emission in H$\beta$, a shell-like profile in H$\gamma$ and several moderate-strength \ion{Fe}{ii} emission lines. This Be nature explains its peculiar position in the CMD, somewhat to the red of the sequence. We classify it as B3\,Ve. Finally, TYC~2148-3261-1 has a similar brightness. We classify it as B5\,V. However, all lines are broad and very shallow. This, together with the very poor \textit{Gaia} DR2 fit (see formal errors in Table~\ref{tabblues}) strongly suggests that it is a binary.

\begin{table*}
        \centering
        \caption{Observed parameters for blue stars with classification spectra. The four stars in the top panel have astrometric parameters compatible with membership. The star in the bottom panel is not an astrometric likely member.}
        \label{tabblues}
        \begin{tabular}{lcccccccc}
                \hline
                \noalign{\smallskip}
Star&  Spectral & pm (RA) & pm (Dec)& $\pi$  & $G$ & $BP$-$RP$  \\ 
&type & (mas) & (mas) & (mas) & (mag) & (mag) \\
\noalign{\smallskip}
\hline
\hline
\noalign{\smallskip}%
LS\,II~$+27^{\circ}$23 &B2.5\,V & $-2.18\pm0.05$ &   $-5.81\pm0.05$ &  $0.42\pm0.03$& 11.74 & 0.71\\
LS\,II~$+27^{\circ}$24  & B2.5\,V &        $-2.02\pm0.05$ &   $-5.91\pm0.05$ &  $0.32\pm0.03$& 11.46 & 0.75 \\
TYC~2148-1327-1 &B3\,Ve&     $-2.07\pm0.04$ &   $-5.86\pm0.05$ &  $0.40\pm0.03$&  12.22 & 0.85\\
TYC 2148-3261-1&B5\,V&$-2.10\pm0.25$ &   $-6.01\pm0.31$ &  $0.47\pm0.16$&  12.13& 0.82 & \\
\noalign{\smallskip}
\hline
\noalign{\smallskip}
LS\,II~$+27^{\circ}$19  & B2.5\,III &        $-2.75\pm0.05$ &   $-5.77\pm0.05$ &  $0.27\pm0.03$& 11.21 & 0.88 \\
\noalign{\smallskip}
                \hline
        \end{tabular}
\end{table*}

\subsection{Cluster parameters}
\label{params}

The stellar content described in the previous section identifies Alicante~13 as a young open cluster. The brightest stars still close to the main sequence have spectral type B2.5\,V. At solar metallicity, this earliest spectral type is typical of clusters in the age range 25\,--\,35~Ma. To derive global parameters, we used $JHK_{{\mathrm S}}$ photometry from the 2MASS catalogue \citep{skru06}. For this, we cross-matched the population selected within 12 arcmin of HD~339063 with the 2MASS point-source catalogue, admitting only objects with well-defined errors in all three bands. Given the high-quality astrometry of both catalogues, we only accepted matches separated by less than 0.3 arcsec. The completeness limit of this catalogue is set at $K_{{\mathrm S}}=14.2$. 

\begin{figure}
   \centering
\resizebox{\columnwidth}{!}{\includegraphics[clip]{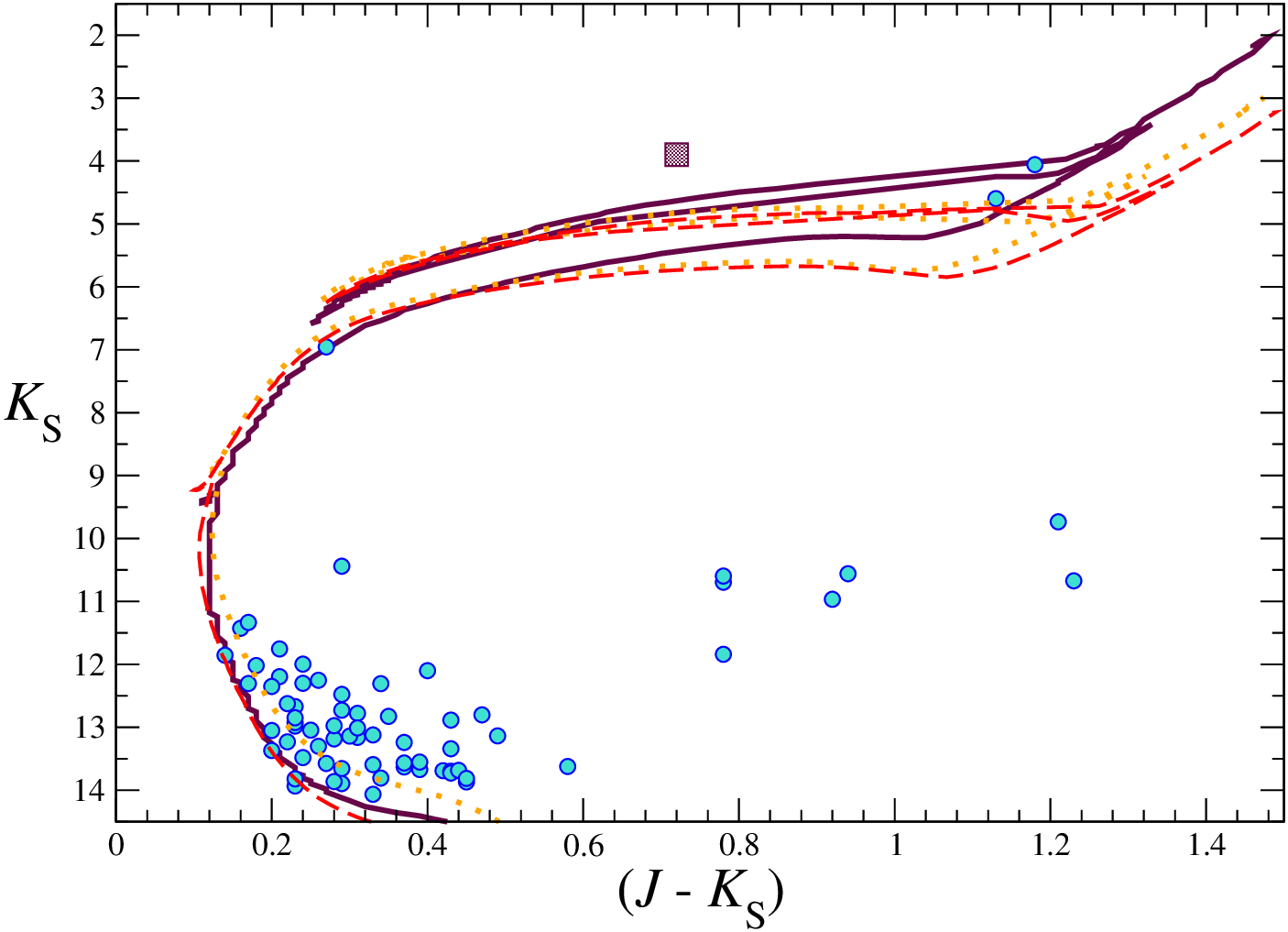}}
   \caption{Infrared CMD for the population of Alicante~13, built with 2MASS data. As with the \textit{Gaia} photometry, there is a small number of astrometric members that do not fit the sequence at $(J-K_{\textrm{S}})\geq0.7$. The thick continuous (maroon) line is a $30\:$Ma isochrone with $DM=12.0$~mag and $A_{V}=1.67$~mag. The dotted (orange) line is a $35\:$Ma isochrone with $DM=11.5$~mag and the same extinction. The dashed (red) line is a $25\:$Ma isochrone with $DM=12.2$~mag. The large square is SV~Vul (as it is saturated in 2MASS, we have taken its average magnitudes from \citealt{fouque07}). The blue stars lying immediately to the right of the main sequence can be considered candidate Be stars.
  \label{ir_cmd}}
    \end{figure}

The mean value of the cluster parallax in DR2 data is 0.37~mas, in total agreement with the DR2 value for SV~Vul itself ($\pi=0.37\pm0.03$~mas). However, there are systematic errors affecting all astrometric parameters in DR2 \citep{luri18}. In particular, there is a zero-point offset \citep{lindegren18}, which seems to depend on colour and perhaps position in the sky \citep{zinn19,khan19}. Given this uncertainty, 
we simply assume a nominal $\pi=0.40$~mas ($d = 2.5\:$kpc), i.e. $DM=12.0$~mag, in agreement with the values obtained by \citet{madore17} for SV~Vul. In Fig.~\ref{ir_cmd}, we plot the 2MASS CMD for the population selected on the basis of their astrometric parameters together with Padova PARSEC isochrones. A 30~Ma isochrone provides a very good fit to the whole blue sequence and the position of the two red supergiants (note that only the red component of BD~$+27^{\circ}$3542  contributes in these bands), with an extinction $A_{V}=1.67$~mag, again fully consistent with the value derived by \citet{madore17} for SV~Vul. In particular, the location of HD~339063 on this diagram is a strong constraint on the fit. Older clusters at shorter distances fail to give a good fit for the upper main sequence. As an example, we show a $35\:$Ma isochrone with $DM=11.5$~mag (the other end of the range suggested for SV~Vul in the literature) and the same extinction. A small decrease or increase of the reddening does not improve the fit, ruling out the older age. 

A younger cluster with the same extinction can provide a good fit to the blue sequence for a slightly higher distance. However, it fails to fit the red supergiants, as the gap in magnitude between the blue and red supergiants increases with age. As an example, in Fig.~\ref{ir_cmd} we also show a $25\:$Ma isochrone with $DM=12.2$~mag.

\begin{figure}
   \centering
\resizebox{\columnwidth}{!}{\includegraphics[clip]{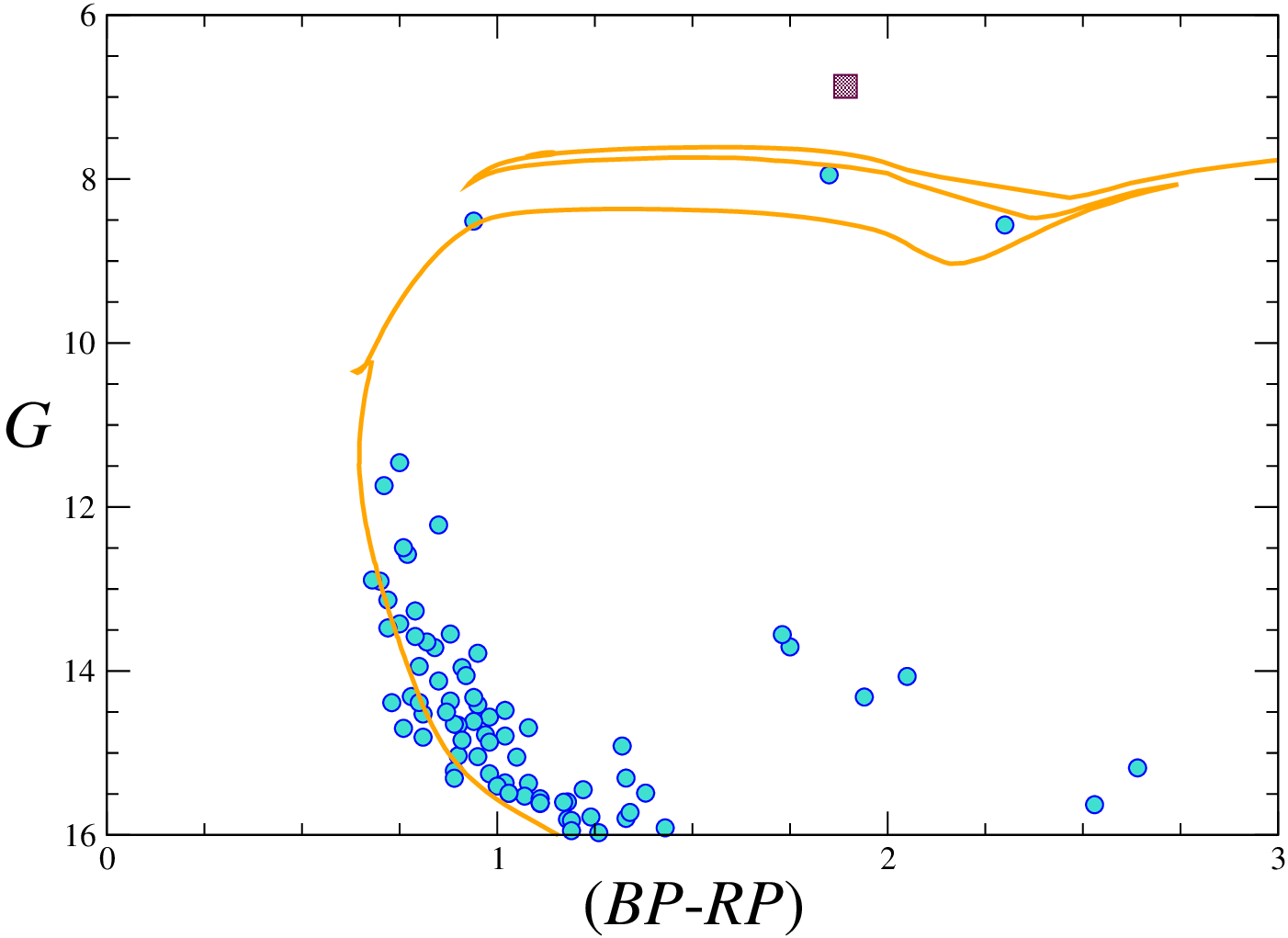}}
   \caption{Best fitting isochrone for the population of Alicante~13, by using \textit{Gaia} photometry. The thick continuous (orange) line is a $35\:$Ma isochrone with $DM=11.7$~mag and $A_{V}=2.0$~mag. The large square is SV~Vul. Different transformations have been used for bright and faint stars, as indicated by \citet{mapw18}.
  \label{gaia_iso}}
    \end{figure}

Despite the excellent fit to the 2MASS CMD, the \textit{Gaia} photometry cannot be fit with the same parameters. In particular, Padova PARSEC isochrones using the passbands of \citet{mapw18} can only fit the sequence with $A_{V}\approx2.0$~mag, a much higher value than implied by the 2MASS data. Given this higher extinction, isochrones give a good fit to the main sequence for a shorter $DM=11.7$~mag. As shown in Fig.~\ref{gaia_iso}, this shorter distance requires a slightly older cluster (35~Ma) to fit the location of the supergiants. The reason for this discrepancy between the extinction derived from optical and infrared photometry is unclear, but it points towards non-standard extinction law. \citet{Turner80_SVul} derives $E(B-V) = 0.65$ and $0.69$ for LS\,II~$+27^{\circ}$23 \& 24, respectively, in agreement with the reddening estimated from the \textit{Gaia} photometry, implying that there are no systematics affecting the optical photometry. Any anomaly in the extinction law, however, will have a smaller impact on the near infrared magnitudes and colours, and thus we favour the age derived from the fit in the 2MASS CMD.

\section{Discussion}

We have identified a previously unknown cluster in the region of Vul OB1, which we name Alicante~13. The long period classical Cepheid SV~Vul is a halo member, located about 9 arcmin West from the centre. About half of the classical Cepheids confirmed as cluster members lie in their respective halos \citep{anderson13}. At a distance of 2.5~kpc, 9 arcmin correspond to 6.5~pc, comparable to the distances of V~Cen, EV~Sct or QZ~Nor to their respective clusters, and much shorter than the distance from V379 Cas to the centre of NGC~129, of which it is an obvious astrometric member \citep[cf.][and \textit{Gaia} DR2 data]{anderson13}.

The 2MASS CMD is best fit by a 30~Ma isochrone, which agrees well with the spectral type of the brightest main-sequence members. According to the isochrone, the mass of the cluster supergiants should be $\sim 9.2\:$M$_{\sun}$. SV~Vul is clearly brighter than the location of the blue loop in the isochrone (about 0.6~mag in $K_{\mathrm{S}}$ and 0.9~mag in $G$). This is not unusual (cf. the position of V340~Nor with respect to the best-fit isochrone for NGC~6067 in \citealt{alonso17}) and could be due to moderately high initial rotation. As an example,  \citet{anderson14} state that a $9\:$M$_{\sun}$ star with high initial rotation will reach the blue loop $\sim5\:$Ma later than a star with zero initial rotation, and so will considerably outlive stars of the same mass with slow rotation. Using isochrones by \citet{georgy13rot}, we find that an $M_{*}=10\:$M$_{\sun}$ star with "standard" initial rotation will traverse the blue loop after $\approx28\:$Ma, reaching a luminosity of $\log L/L_{\sun} = 4.3$. For SV~Vul, the $P$/$L$ relation of \citet{anderson13} gives $M_{V}=-5.8$, while that of \citet{madore17} gives $M_{V}=-5.7$. Assuming a $BC$ not very different from zero for an "average" spectral type around G0\,Iab, this means $\log L/L_{\sun} \approx 4.2$. On the other hand, a star of $M_{*}=9.2\:$M$_{\sun}$ with low initial rotation will be leaving the loop (i.e. becoming a red supergiant) after $30\:$Ma, at a $\log L/L_{\sun} = 3.9$. Therefore the timescales and luminosities agree very well with the fit to the rest of the cluster obtained with a PARSEC isochrone. Interestingly, according to \citet{anderson14}, fast rotators with masses higher than $10\:$M$_{\sun}$ do not experience blue loops. Therefore SV~Vul would be very close to the highest mass for which a fast rotator may appear as a Cepheid variable\footnote{We can also speculate that SV~Vul is brighter than the other supergiants because it is a mass gainer in a binary interaction, which resulted in a higher mass than those of other members. However, it is unlikely that mass gain would have resulted in a slow rotator, while the difference in magnitude does not allow for a very large mass difference. Therefore the hypothesis of a somewhat more massive fast rotator seems to fit better.}.

\begin{figure}
   \centering
\resizebox{\columnwidth}{!}{\includegraphics[clip]{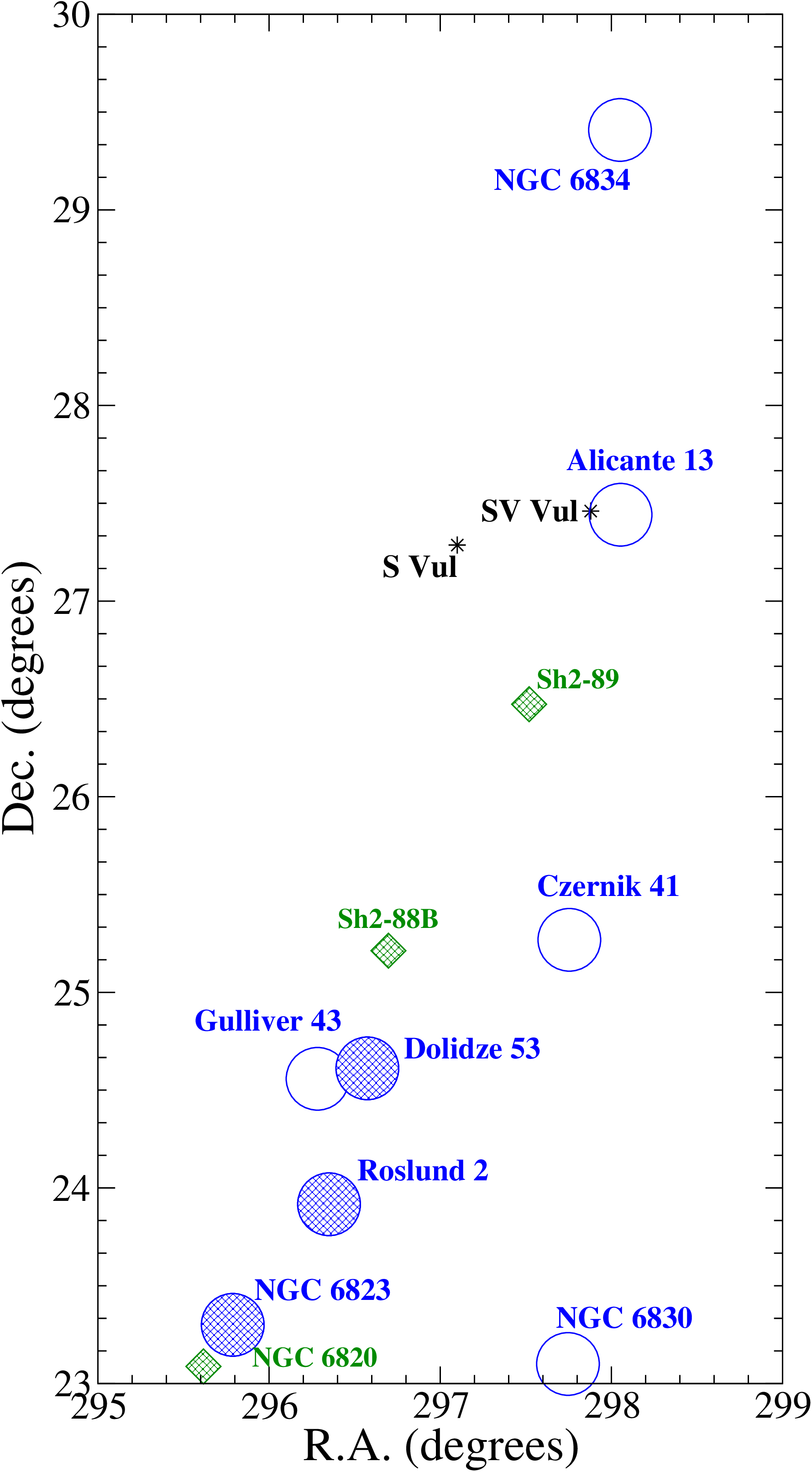}}
   \caption{Relative position in the sky for objects that could be related to Vul~OB1. Circles indicate open clusters, while diamonds mark \ion{H}{ii} regions. All filled symbols indicate objects associated with star-forming nebulosities and thus younger than a few Ma. 
  \label{scheme}}
    \end{figure}

Membership in a cluster allows a good estimation of the mass of SV~Vul. In the past, it had been considered a field member of the Vul~OB1 association, but this region of the sky is very complex; three different OB associations are believed to be projected one on top of the other, Vul~OB4 at $\sim1\:$kpc, Vul~OB1 at $\sim2\:$kpc and Vul~OB2 at $\sim4\:$kpc \citep{Turner80_SVul}. Figure~\ref{scheme} shows the relative positions of relevant objects in this area. The star forming cluster NGC 6823, generally identified as the core of Vul~OB1 lies about four degrees South of SV~Vul and has \textit{Gaia} DR2 astrometric values pmRA =$-1.7\:$mas/yr, pmDec =$-5.3\:$mas/yr, $\pi=0.45\:$mas \citep{cantat18}. The few stars visible in the vicinity of the nearby \ion{H}{ii} region NGC~6820 have compatible values. The nearby cluster NGC~6830, which is much older ($\sim 100\:$Ma), has pmRA =$-3.2\:$mas/yr, pmDec =$-5.8\:$mas/yr, $\pi=0.42\:$mas. More to the North, Roslund~2 has pmRA =$-1.7\:$mas/yr, pmDec =$-5.1\:$mas/yr, $\pi=0.46\:$mas, fully compatible with the values for NGC~6823. Roslund 2 is also a very young cluster, although somewhat older than NGC~6823, as it contains the blue supergiant HD~186745 (B8\,Ia).

As we move further to the North, we find the small star-forming cluster Dolidze~53 (pmRA =$-1.5\:$mas/yr, pmDec =$-4.6\:$mas/yr, $\pi=0.48\:$mas), and a number of early-type stars associated with the \ion{H}{ii} region Sh2-S88B, such as HD 338926 (pmRA =$-2.05\pm0.06\:$mas/yr, pmDec =$-3.37\pm0.06\:$mas/yr, $\pi=0.45\pm0.04\:$mas) or HD 338916 (pmRA =$-3.68\pm0.06\:$mas/yr, pmDec =$-3.39\pm0.05\:$mas/yr, $\pi=0.49\pm0.04\:$mas), whose parallaxes are fully compatible with those of the association, but presenting very different proper motions. Based on their radial velocities, both Sh2-S88B and Sh2-89 are considered members of Vul~OB1, conforming a large star-forming complex \citep[e.g.][]{turner86}. The parallax to Alicante~13 ($\pi=0.37\:$mas) is not sufficiently far away from the value for this complex to completely rule out an association (especially given the possibility of important systematics affecting sources that are more than four degrees away), but the cluster is considerably older. Two nearby clusters have very similar astrometric parameters in the catalogue of \citet{cantat18}, Czernik~41 (pmRA =$-2.9\:$mas/yr, pmDec =$-6.2\:$mas/yr, $\pi=0.37\:$mas) and Gulliver~34  (pmRA =$-2.9\:$mas/yr, pmDec =$-5.8\:$mas/yr, $\pi=0.35\:$mas). Both are dispersed, little-studied clusters, without reliable estimates of their ages.

\begin{figure}
   \centering
\resizebox{\columnwidth}{!}{\includegraphics[angle=-90, clip]{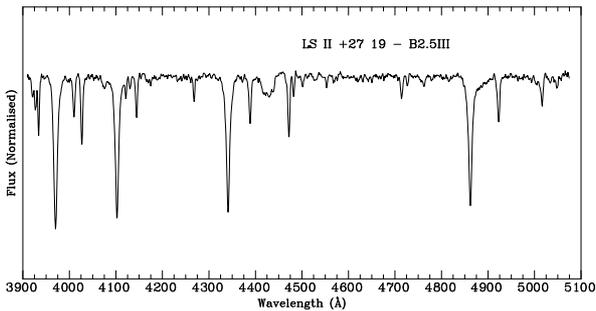}}
   \caption{Classification spectrum of a background star in the field of Alicante~13.
  \label{intruder}}
    \end{figure}

The parameters for NGC~6834 (pmRA =$-2.5\:$mas/yr, pmDec =$-5.1\:$mas/yr, $\pi=0.27\:$mas), on the other hand, cannot be reconciled with those of the the Vul OB1 complex, against previous suggestions. It seems to be rather more distant. The \textit{Gaia} parameters for S~Vul, assumed in the past to be a member of Vul~OB2 \citep{Turner80_SVul}, are pmRA =$-3.40\pm0.06\:$mas/yr, pmDec =$-5.92\pm0.06\:$mas/yr, $\pi=0.31\pm0.04\:$mas (in agreement with the HST parallax  $\pi=0.32\pm0.04\:$mas; \citealt{riess18a}). In view of this, it would seem that the classical view of two separate associations, Vul~OB1 and Vul~OB2, projected over the same region, but with clearly distinct distances of $\sim2$ and $\sim4$~kpc \citep{Turner80_SVul} has to be abandoned\footnote{Note that, with the current \textit{Gaia} DR2 data, the parallaxes for SV~Vul and S~Vul are consistent within their errors.}. The emerging picture is much more complex, with objects projected over a wide range of distances. As an example, within the central region of Alicante~13, we find the catalogued OB star LS\,II~$+27^{\circ}$19, with DR2 parameters pmRA =$-2.75\pm0.05\:$mas/yr, pmDec =$-5.77\pm0.05\:$mas/yr, $\pi=0.27\pm0.03\:$mas. As can be seen in Fig.~\ref{intruder} and Table~\ref{tabblues}, this is a giant star of the same spectral type as those at the top of the cluster sequence. Its parallax is inconsistent with that of the cluster at $>2\,\sigma$ and it is only very slightly brighter than the main sequence stars, indicating a longer distance modulus. Its reddening, however, is only very slightly higher. Using the $UBV$ photometry from \citet{Turner80_SVul} and our spectral type, we calculate a $DM = 12.8$ ($d=3.6\:$~kpc), in good agreement with its parallax\footnote{Noting that we assume a standard reddening law, which, in view of the discrepancy between fits to the optical and infrared photometry, does not seem to hold.}. All this suggests that, beyond the obscuring clouds associated to Vul~OB1, at about 2~kpc, the reddening increases very little in this direction until at least close to 4~kpc, resulting in the projection of large numbers of stars at different distances on to what seems to be a single association. 
With the current \textit{Gaia} DR2, this hypothesis cannot be tested. As long as systematic errors as high as 0.1~mas are possible \citep{luri18}, differentiating between 2 and 4~kpc is not straightforward. Despite this, it is worth noting that the \textit{Gaia} parallaxes for the two Cepheids (SV~Vul and S~Vul), which are complicated \textit{Gaia} targets, are fully consistent with previous estimates (including the HST parallax for S~Vul). At the same time, stars in their neighbourhoods that were classified as members of Vul~OB2 by \citet{Turner80_SVul} have parallaxes ranging from $\pi=0.42\pm0.03$ (LS\,II~$+27^{\circ}$23) to $0.15\pm0.04$ (HD~332848), which strongly suggests that they are not all at the same distance. Further \textit{Gaia} releases will clarify this issue. 

Whatever the case, it is clear that SV~Vul is not directly linked to the star-forming clusters in Vul~OB1, but to a rather older population. Its membership in Alicante~13 shows that one of the most luminous Cepheid variables in the Milky Way has a mass of only $\approx10\:$M$_{\sun}$. Indeed, there is no strong reason to believe that the even more luminous S~Vul belongs to a younger population. This is in agreement with the models presented by \citet{anderson14}, where the highest mass for a star with typical rotation (their "average" rotation rate of $\omega = 0.5$) to experience a blue loop is slightly higher than $10\:$M$_{\sun}$. Models with zero rotation allow Cepheids as massive as $11.5\:$M$_{\sun}$. Such null rotation is in fact unphysical, but stars with low rotational velocities do indeed exist, and they could become Cepheids for masses of at least $11\:$M$_{\sun}$, although with a luminosity comparable to that of the most luminous "rotating" Cepheids ($\log L/L_{\sun} \approx4.3$), considered by \citet{anderson14} the upper limit in luminosity for Cepheids.

Among Cepheids in clusters, the only strong evidence for a mass higher than $\approx10\:$M$_{\sun}$ is given by the 23~d Cepheid in vdBH~222 \citep{clark15}. At the distance implied by the Cepheid properties ($\approx6$~kpc), the cluster has an age $\approx20\:$Ma \citep{marco14} and thus the properties of this Cepheid, which has a luminosity similar to that of SV~Vul can be explained by a $\ga11\:$M$_{\sun}$ star with low rotation. For example, using isochrones by \citet{georgy13rot}, we find that an $M_{*}=11.2\:$M$_{\sun}$ star with low rotation ($\omega = 0.1$) will be in the blue loop at an age of $\approx20.3\:$Ma, with $\log L/L_{\sun} \approx4.3$. Again, we find very good agreement, even if we have to consider that all tracks by \citet{georgy13rot} between 9 and~$12\:$M$_{\sun}$ are interpolated, and not directly calculated. Although further study of vdBH~222 is needed for a better characterisation, all observations of Milky Way Cepheids in clusters seem to agree to a very good degree with the models of \citet{anderson14}. Further \textit{Gaia} data releases will undoubtedly provide us with a larger number of high-luminosity Cepheids with accurate distances.

Meanwhile, the very accurate proper motions provided by DR2 are already allowing us to discover new clusters with very low background contrast, such as Alicante~13, which would be completely inconspicuous except for the fortunate presence of three (super)giants within one arcmin. With an accurate distance to the cluster, successive \textit{Gaia} releases may turn SV~Vul into a key landmark for the local distance scale and the study of the evolutionary status of long-period Cepheids. 

\section{Conclusions}
We find that the 45~d classical Cepheid SV~Vul is a certain astrometric halo member of a new young open cluster, Alicante~13. The star is slightly brighter than predicted by the best fit isochrones, suggesting that it started its life as a fast rotator. With the best fit age, stars close to the main sequence have $\approx9\:$M$_{\sun}$ and SV~Vul has a mass of $\approx10\:$M$_{\sun}$, in agreement with the highest mass compatible with a fast rotator entering the blue loop \citep{anderson14}. Isochrone fit to the cluster CMD favours a distance of 2.5~kpc, in good agreement with its \textit{Gaia} parallax and the distance proposed by \citet{madore17}. There is, however, a discrepancy between the amount of extinction needed to fit the optical and the infrared photometry, suggestive of a moderately non-standard reddening law. The connection of Alicante~13 to the Vul~OB1 association, dominated by NGC~6823, is unclear. In fact, only clusters with current star formation seem to be certain members of the association. With upcoming \textit{Gaia} releases providing a definitive distance, SV~Vul will become an anchoring point in the study of long-period classical Cepheids.

\section*{Note added in proof}
After submission of this paper, \citet{castrogin20} presented the results of a systematic search for clusters, conducted by applying a machine learning methodology to \textit{Gaia} DR2 data. Among their 245 class A candidate clusters, they find Alicante~13, under the name UBC~130, with centre at RA: 19:52:12.17, Dec: +27:26:42.9. Their estimated parameters are $\pi = 0.39\pm0.03\:$mas, pmRA=$-2.11\pm0.08\:$mas/yr,  pmDec =$-5.85\pm0.08\:$mas/yr, fully consistent with the values derived in this paper. According to their criteria, TYC~2148-893-1 is an outlying cluster member. If so, it must be a third red supergiant. Its \textit{Gaia} DR2 radial velocity  $0.86\pm0.24\:$km\,s$^{-1}$ is compatible with those of other members. However, both its \textit{Gaia} and 2MASS magnitudes, $K_{\mathrm{S}}=5.44$, $(J-K_{\mathrm{S}})=1.21$, place it below the cluster isochrone. Spectroscopic data will be needed to assess its membership.

\section*{Acknowledgements}

We thank the anonymous referee for their prompt response and encouraging comments.
This research is partially supported by the Spanish Government under grants AYA2015-68012-C2-2-P and PGC2018-093741-B-C21/C22 (MICIU/AEI/FEDER, UE).

The Nordic Optical Telescope is operated by the Nordic Optical Telescope Scientific Association at the Observatorio del Roque de los Muchachos (La Palma, Spain)
of the Instituto de Astrof\'isica de Canarias. Observations have been taken as part of the Spanish service time, and we gratefully acknowledge the work of the support astronomers. The Starlink software (Currie et al.\ 2014) is currently supported by the East Asian Observatory. This work has made use of data from the European Space Agency (ESA) mission
{\it Gaia} (\url{https://www.cosmos.esa.int/gaia}), processed by the {\it Gaia}
Data Processing and Analysis Consortium (DPAC,
\url{https://www.cosmos.esa.int/web/gaia/dpac/consortium}). Funding for the DPAC
has been provided by national institutions, in particular the institutions
participating in the {\it Gaia} Multilateral Agreement.

 This research has made use of the Simbad, Vizier and Aladin services developed at the Centre de Donn\'ees Astronomiques de Strasbourg, France. It also makes use of data products from 
the Two Micron All Sky Survey, which is a joint project of the University of
Massachusetts and the Infrared Processing and Analysis
Center/California Institute of Technology, funded by the National
Aeronautics and Space Administration and the National Science
Foundation. 



\bibliographystyle{mnras}
\bibliography{clusters,rsgs,bins,gaia,obstars,cepheid} 





\bsp	
\label{lastpage}
\end{document}